\documentclass[sigplan,10pt,nonacm]{acmart}

\acmSubmissionID{480}
\acmConference[EuroSys '27]{European Conference on Computer Systems}{April 19--23, 2027}{Rabat, Morocco}

\renewcommand\footnotetextcopyrightpermission[1]{}
\setcopyright{none}
\usepackage{enumitem}
\usepackage{xspace}
\usepackage{subcaption}
\usepackage{tikz}
\usepackage{pifont}

\setlist[itemize]{leftmargin=*}
\setlist[enumerate]{leftmargin=*}

\newcommand{\tinyskip}{\par\addvspace{2pt}}
\newcommand{\mypar}[1]{\tinyskip\noindent\textbf{#1.}\xspace}

\newcommand*\myc[1]{%
  \scalebox{0.78}{%
    \begin{tikzpicture}[baseline=-3pt]
      \node[draw,circle,inner sep=0.5pt,fill=black]
        {\textcolor{white}{\textsf{\textbf{#1}}}};
    \end{tikzpicture}%
  }%
}

\newcommand{\cmark}{\ding{51}}
\newcommand{\xmark}{\ding{55}}

\newcommand{\sys}{Chase\xspace}

\makeatletter
\def\@authorfont{\large\normalfont}
\makeatother

\begin{document}

\title{\sys: Arbitrage-Aware Scheduling for Geo-Distributed AI Infrastructure}

\author{Matej Sandor\textsuperscript{$*\dagger$}, Yao Fu\textsuperscript{$*$}, Chi Xing, Yinsicheng Jiang, Guanhao Su, Xuan Sun, Le Xu, Adam Barker, Luo Mai}
\affiliation{%
  \institution{University of Edinburgh, UK}
  \country{}
}
\thanks{\textsuperscript{$*$}Equal contribution.\newline \textsuperscript{$\dagger$}Correspondence to: \href{mailto:matej.sandor@ed.ac.uk}{matej.sandor@ed.ac.uk}}
\renewcommand{\shortauthors}{Sandor and Fu, et al.}

\begin{abstract}
AI infrastructure has become one of the fastest-growing consumers of electricity, yet GPU workloads are still scheduled with little regard for when and where energy is cheap. Electricity prices vary widely across regions and time. A provider whose data centers span several markets can therefore run the same work at very different costs, but only if the scheduler plans across regions and time horizons. Existing schedulers target a single region, a single workload type, or a single control horizon, so they capture only part of this opportunity. We present Chase, a job-agnostic scheduler for geo-distributed AI infrastructure. Chase represents every workload as a set of pre-validated execution options and plans across time horizons, combining long-term price trends with near-real-time refinement. On a GH200 cluster and in emulation with real price traces, Chase reduces energy cost by 27–35\% on volatile markets and turns compute spending into net revenue when prices go negative.
\end{abstract}

\maketitle

\section{Introduction}

\begin{figure}[t]
  \centering
  \includegraphics[width=\linewidth]{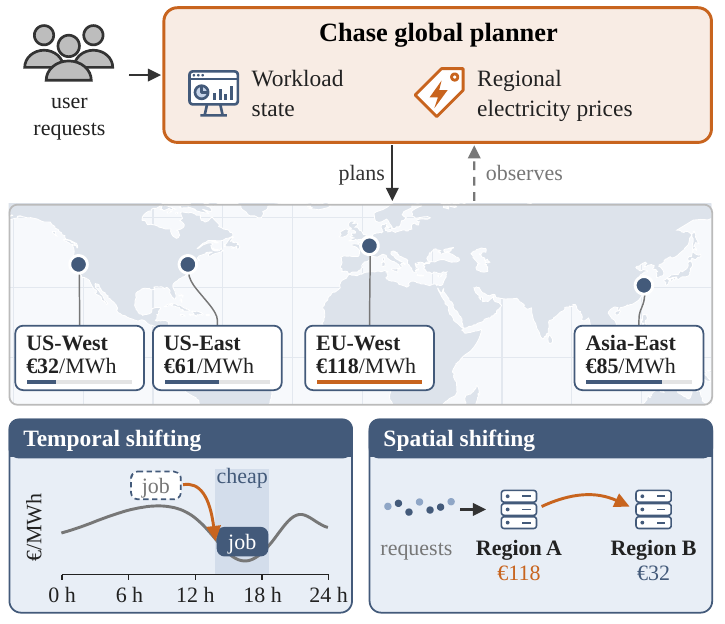}
  \caption{Overview of \textbf{\sys}. A global planning layer observes each region's electricity price and cluster state and places jobs. Work moves along two axes: in time, to a cheaper hour in the same region, and in space, to a cheaper region at the same hour.}
  \label{fig:energy-view}
\end{figure}

AI infrastructure is one of the fastest-growing consumers of electricity~\cite{shehabi2024lbnl, iea2025energy}. Foundation-model training~\cite{strubell2019energy, patterson2021carbon, hu2024characterization} and large-scale serving~\cite{patel2024power, qiu2024muserve, stojkovic2025tapas} increasingly run on GPU fleets distributed across data centers worldwide~\cite{yuan2022decentralized, strati2025sailor, choudhury2024mast, mao2025skyserve, xia2025skywalker}. At this scale, electricity is both a substantial operating cost and a cost that varies significantly across regions and time~\cite{qureshi2009cutting}. Increasing renewable generation further amplifies this variation: when renewable supply exceeds demand, limited demand can drive wholesale prices very low or even negative~\cite{sensfuss2008merit, acun2023carbon, radovanovic2023carbon}. The price variation creates an opportunity for \emph{cross-region energy arbitrage}: shifting computation to locations and times with lower electricity prices.

Exploiting energy arbitrage requires coordinating two forms of flexibility, shown in Figure~\ref{fig:energy-view}. Computation can move \emph{in time}, by pausing during expensive periods and resuming when prices fall, or \emph{in space}, by moving to a region with cheaper electricity. Different workloads expose different forms of flexibility. Batch jobs can often be delayed or paused subject to a completion deadline~\cite{shukla2022singularity, strati2025sailor}, whereas latency-critical services must remain available and satisfy latency targets but can often be placed or routed across regions~\cite{mao2025skyserve, xia2025skywalker}. Cheap capacity is also limited, causing heterogeneous workloads to compete for the same low-price periods and regions. Energy-aware scheduling therefore requires joint allocation across workloads, regions, and time.

The allocation problem also spans multiple time scales. Electricity prices have predictable structure over a horizon of days: day-ahead markets clear prices for the following 24 hours~\cite{entsoe2026transparency} and forecasts extend that view further~\cite{lago2021forecasting}, so flexible workloads can be positioned around recurring low-price periods. Prices and available cluster capacity, however, also change on much shorter time scales. A schedule constructed from long-term forecasts can become suboptimal before execution, while a scheduler operating only at short horizons lacks sufficient lookahead to exploit predictable price structure.

Existing systems only address individual aspects of the problem. Multi-cluster schedulers coordinate workloads across geographically distributed clusters, but do not treat electricity price as a first-class scheduling signal~\cite{choudhury2024mast, shukla2022singularity}. Energy- and carbon-aware systems adapt workload execution to energy signals, but generally target particular workload types or scheduling settings~\cite{hanafy2023carbonscaler, gsteiger2024caribou, xu2025green}. Geographically distributed serving systems exploit placement and routing flexibility, but focus on serving objectives rather than jointly optimizing electricity cost across regions~\cite{mao2025skyserve, xia2025skywalker}. None of these systems plan at more than one scheduling horizon, and hence fail to combine long-term price structure with short-lived changes in price and capacity.

Our key insight is that low-price execution windows yield the greatest savings when assigned to jobs with higher energy consumption per unit of work, while preserving each job's resource and performance requirements. Different jobs are comparable on power usage, so the scheduler can distribute resources without understanding execution details. The scheduler only needs a compact description of \emph{where} a workload can run, \emph{what} each placement costs and delivers, and \emph{which constraints} it must satisfy, while existing cluster schedulers and job-level engines retain responsibility for execution.

\sys{} realizes this insight as a scheduler. Every job enters as a set of pre-validated execution options, each a resource configuration with its profiled throughput and power draw, together with a specification of what a valid schedule must satisfy. A long-horizon planner distributes each job's work based on multi-day price forecasts at coarse resolution, and a short-horizon refiner binds that work to an hour, a region, and an execution option using the prices, free capacity, and job progress it observes. Regional cluster schedulers and job-level engines carry out the resulting schedule, so \sys{} requires no change to how a job executes.
Specifically, this paper makes the following contributions:

\mypar{(1) Job scheduling abstraction}
We introduce a workload-independent interface in which each job is a set of pre-validated execution options together with lightweight requirements, such as a deadline or an SLO. It exposes what global placement needs while hiding workload-specific execution detail.

\mypar{(2) Multi-horizon scheduling}
We develop a two-level scheduler that captures predictable price structure while retaining the ability to exploit short-lived changes in price and capacity.

\mypar{(3) \sys{}, a tractable cross-region planner}
We show that joint optimization over workloads, regions, and time remains tractable at provider scale: pre-validated options, option pruning, and the horizon split bound the problem the solver sees, so cost grows with the number of scheduling decisions rather than with the size of the GPU fleet.

\mypar{(4) Extensive evaluation on real hardware and in emulation}
We evaluate \sys{} on a GH200 cluster using three real electricity-price traces with training and inference workloads. On volatile traces, \sys{} cuts energy cost by 27--35\%, against at most 20\% for the strongest of GREEN and Caribou. A solver stress test shows that Chase's formulation remains tractable even at provider scale: the fine-grained placement problem for 200 jobs across 39 regions and over 40,000 GPUs solves in under 25 seconds, well within the hourly re-solve budget.

\section{Background and Motivation}
\label{sec:background}

AI infrastructure is increasingly distributed across geographically separated GPU clusters, creating flexibility in where and when workloads execute. At the same time, electricity prices vary substantially between regions and over time, creating opportunities to reduce electricity cost through shifting workloads. In this section, we characterize the workload flexibility and price variation that enable this opportunity and derive scheduling requirements addressed by \sys. 

\subsection{Geo-Distributed AI Infrastructure and Workloads}
\label{sec:geo}

Modern AI infrastructure is inherently geo-distributed. Major cloud providers and AI companies operate GPU clusters across continents~\cite{aws2026regions, azure2026regions}, driven by capacity constraints and by the need to serve users worldwide, and a single organization may run pre-training in one region, fine-tuning in another, and serve inference traffic from several more. This geo-distribution creates an opportunity to exploit regional differences in electricity prices when deciding where and when workloads execute. 

A scheduler that shifts work across regions must account for three workload properties. \textbf{Preemptibility} determines the cost of interrupting and resuming execution; \textbf{migration cost} captures the cost of moving job state between regions; and \textbf{job requirements} capture constraints such as deadlines, availability, and latency. We therefore distinguish two broad classes of AI workloads:

\mypar{Deadline-driven workloads}
Pre-training~\cite{grattafiori2024llama3, jiang2024megascale} and post-training~\cite{ouyang2022instructgpt, shao2024deepseekmath, sheng2025hybridflow} are iterative, deadline-driven workloads whose deadlines often provide substantial scheduling slack. These workloads already use checkpointing for fault tolerance~\cite{mohan2021checkfreq, wan2025bytecheckpoint, kokolis2025reliability}, so pause and resume are well established. They can therefore be interrupted and relocated, at a cost set mainly by the model and optimizer state that must be transferred~\cite{zhao2023fsdp}. SkyPilot~\cite{yang2023skypilot} and Skyplane~\cite{jain2023skyplane} demonstrate cross-region state transfer and its overhead, and preemptible-capacity systems restore execution state after eviction~\cite{thorpe2023bamboo, duan2024parcae}.

\mypar{Latency-critical workloads}
In contrast, online serving workloads must maintain availability while satisfying latency and throughput objectives~\cite{zhong2024distserve}. They cannot tolerate downtime, but support migration as long as service continuity is preserved. ServerlessLLM~\cite{fu2024serverlessllm} enables fast model loading to support migration with limited interruption, while Llumnix~\cite{sun2024llumnix} migrates live requests across instances. Cross-region placement adds network round-trip latency, but for large models, request processing dominates this overhead~\cite{mao2025skyserve, xia2025skywalker}.

\subsection{Electricity Price as a Scheduling Objective}
\label{sec:why-price}

\begin{figure*}[t]
  \centering
  \includegraphics[width=\textwidth]{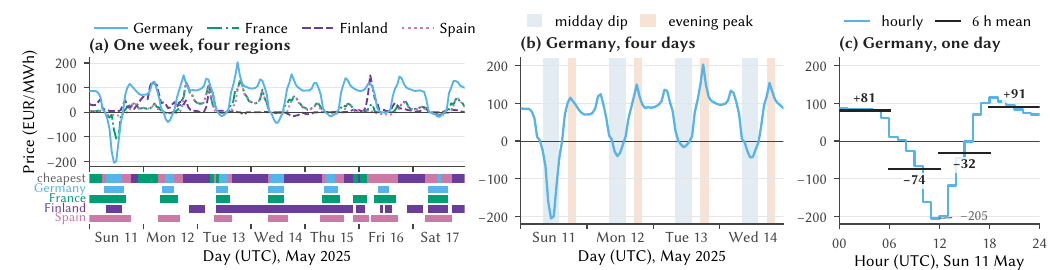}
  \Description{Three panels. The left panel plots hourly day-ahead prices for Germany, France, Finland, and Spain over one week, above five strips that mark the cheapest region each hour and the hours each region prices below ten euros per megawatt-hour. The middle panel plots Germany over four days with the midday dip and evening peak of each day shaded. The right panel plots Germany over one day with four six-hour period averages drawn as steps.}
  \caption{European day-ahead prices over the week of 11 May 2025, UTC (ENTSO-E~\cite{ember2026electricity, entsoe2026transparency}). The strips under (a) give the cheapest region of each hour and the hours each region prices below 10~EUR/MWh. The shaded bands in (b) are the same two hours of every day. The steps in (c) are the four 6~h period averages of that panel's hourly series, each labeled in EUR/MWh; $-$205 marks the lowest single hour.}
  \label{fig:price-traces}
\end{figure*}

Scheduling against electricity prices is known to reduce cost~\cite{qureshi2009cutting}, and the growing power demand of AI infrastructure makes it increasingly relevant~\cite{shehabi2024lbnl, iea2025energy, epri2024powering}. Training Llama~3.1 405B used 30.84M H100 GPU-hours~\cite{meta2024llama31, nvidia2026h100}, about 21.6~GWh at the 700~W TDP, so each 100~EUR/MWh of price difference is worth 2.16M~EUR, and the prices in Figure~\ref{fig:price-traces} span over 400~EUR/MWh. Price is also a useful, though imperfect, proxy for renewable availability~\cite{sukprasert2024limitations}, since low and negative prices often occur when renewable generation exceeds demand~\cite{sensfuss2008merit, fanone2013negative, seel2021negative}.

Energy use also matters beyond cost. The same run emitted an estimated 8{,}930~tCO$_2$eq~\cite{meta2024llama31}, and operators face growing carbon constraints, from net-zero commitments~\cite{acun2023carbon} to the EU requirement that data centers above 500~kW report their energy use and renewable share annually~\cite{eu2023eed}. Shifting work toward renewable-rich periods and regions reduces emissions even where price differences are small~\cite{radovanovic2023carbon}.

\sys{} therefore treats performance requirements as hard constraints and energy cost as the scheduling objective, choosing \emph{when} and \emph{where} work runs. Each configuration reduces to two profiled quantities, FLOP throughput and energy consumption, so one formulation minimizes either cost or carbon through the per-kWh price or carbon intensity (\S\ref{sec:multi-horizon}). Deadline-driven jobs specify a total FLOP requirement, and latency-critical jobs specify a required FLOP rate. This complements techniques that reduce workload energy~\cite{you2023zeus, chung2024perseus, tschand2025mlperfpower}, which improve a configuration's efficiency without changing the scheduling objective.

\subsection{Energy Arbitrage Opportunities}

We analyze electricity prices and derive three observations.

\mypar{Insight 1: Cheap capacity is limited}
A low-price period provides a finite amount of cheap compute capacity, so the scheduler must determine which of the competing jobs benefits most from occupying it. That depends on a job's flexibility and on its energy consumption. Across the workloads in our evaluation, we measure a 1.6$\times$ difference in power per GPU, consistent with the difference between training and serving clusters reported in production~\cite{patel2024power}. Assigning a high-power job to a low-price period can therefore yield substantially greater savings than assigning the same capacity to a low-power job, which motivates joint scheduling across jobs and the job agnostic abstraction.

\mypar{Insight 2: Temporal and regional price differences create sustained scheduling opportunities}
Electricity prices vary substantially across both time and regions~\cite{souza2023ecovisor, sun2024skybox, naug2024sustaindc, guillenperez2025dcclusteropt}. Figure~\ref{fig:price-traces}(a) shows one week of day-ahead prices across four European regions~\cite{entsoe2026transparency}. Across the traces, 314 region-hours fall below 10~EUR/MWh, 58\% of them negative, with low-price periods lasting 9.8 hours on average. In Finland, for example, prices remain below 10~EUR/MWh for roughly two and a half days, creating substantial temporal shifting opportunity. 

Regional variation provides a second opportunity: the cheapest region changes 31 times over the week, with prices ranging from $-205$ to over $+200$~EUR/MWh. Because regional prices are not perfectly correlated~\cite{trebbien2024patterns}, a scheduler confined to one region can exploit only part of the low-price capacity.

\mypar{Insight 3: Opportunities emerge at multiple time horizons}
Opportunities emerge at different time scales. Some are predictable days ahead, such as daily cycles and extended low-price periods, while others become visible only near execution as forecasts and cluster state improve~\cite{lago2021forecasting}. This creates a granularity tradeoff: long-horizon planning can exploit predictable structure but may become stale, while short-horizon planning can react to changes but miss opportunities that require advance positioning. Figure~\ref{fig:price-traces}(b) illustrates the former: Germany shows a recurring midday dip and evening peak across four days, visible even at six-hour granularity. Figure~\ref{fig:price-traces}(c) illustrates the latter: within one six-hour period, hourly prices fall well below the average. Effective scheduling thus requires both coarse long-horizon planning and fine-grained short-horizon refinement. Existing systems only plan at one scheduling horizon.

\begin{table}[t]
\centering
\caption{Scheduler comparison of existing systems to \sys.}
\label{tab:related-comparison}
\small
\setlength{\tabcolsep}{2.5pt}
\begin{tabular}{@{}lccccc@{}}
\toprule
 & Multi- & Temporal & Spatial & Job & Multi- \\
System & region & shifting & shifting & agnostic & horizon \\
\midrule
MAST~\cite{choudhury2024mast}            & \cmark & \xmark & \xmark & \xmark & \xmark \\
SkyServe~\cite{mao2025skyserve}          & \cmark & \xmark & \cmark & \xmark & \xmark \\
GREEN~\cite{xu2025green}                 & \xmark & \cmark & \xmark & \xmark & \xmark \\
GreenFlow~\cite{gu2025greenflow}         & \xmark & \cmark & \xmark & \xmark & \xmark \\
Caribou~\cite{gsteiger2024caribou}       & \cmark & \xmark & \cmark & \xmark & \xmark \\
\midrule
\sys{}                                   & \cmark & \cmark & \cmark & \cmark & \cmark \\
\bottomrule
\end{tabular}
\end{table}

\mypar{The need for a unified scheduler}
Together, these observations form the core features required to capture the full opportunity, and Table~\ref{tab:related-comparison} shows that prior work addresses only subsets of them. They also cannot be optimized independently: low-price GPU-hours are a shared, scarce resource, so scheduling each workload separately can consume capacity another workload could use better. \sys{} coordinates decisions across workloads, time, and regions.

\section{Arbitrage-Aware Scheduling}
\label{sec:model}

\begin{figure}[t]
\centering
\includegraphics[width=\columnwidth]{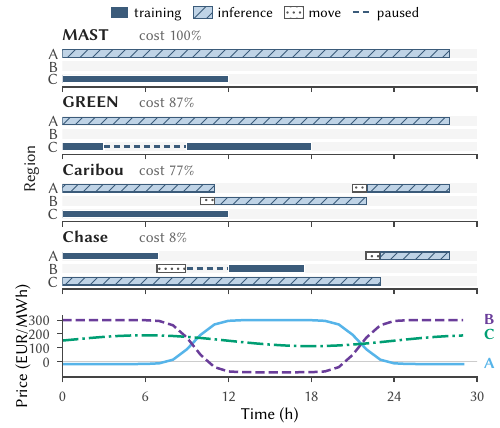}
\caption{A post-training job with a 24~h deadline and an inference service scheduled over 30~h across three regions, each with capacity for one job. Static placement commits both jobs at submission; GREEN~\cite{xu2025green} shifts the post-training job in time but not across regions; Caribou~\cite{gsteiger2024caribou} enables migration but optimizes workloads independently; \sys{} jointly allocates low-price capacity across both workloads. Costs are normalized to static placement.}
\label{fig:scheduling_comparison}
\end{figure}

Figure~\ref{fig:scheduling_comparison} illustrates the opportunity for global scheduling. We consider a 12-hour post-training job with a 24-hour deadline and an inference service running across three regions, each with capacity for one job. The post-training job can pause and migrate, while the service must remain available but can migrate. Prices differ substantially across regions and time, and the post-training job consumes more power, making scarce low-price capacity particularly valuable to it.

MAST~\cite{choudhury2024mast}, our static-placement baseline, commits each job at submission, so arrival order determines who receives the cheapest capacity. GREEN~\cite{xu2025green} adds temporal shifting but stays in one region, while Caribou~\cite{gsteiger2024caribou} adds migration but optimizes workloads independently. In this example, they reduce cost by 13\% and 23\%, respectively. \sys{} instead jointly schedules both workloads, assigning the low-price periods to the higher-power job and reducing cost by 92\% relative to static placement.

Composing the baselines still leaves 69\% between Caribou and \sys{}, because neither coordinates scarce low-price capacity across workloads. Our ablation confirms this: migration alone saves 27\%, while joint scheduling reaches 54\% (\S\ref{sec:eval:horizon}). Closing this gap presents three challenges:

\mypar{C1. Comparing heterogeneous jobs for the same capacity}
Heterogeneous workloads differ in execution mechanisms, parallelism, hardware requirements, and service constraints, yet compete for the same regional capacity and low-price periods. The scheduler therefore needs a common representation that makes different jobs comparable without encoding workload-specific execution semantics.

\mypar{C2. Coordinating scheduling decisions across time horizons}
No single horizon captures the full opportunity. Long-horizon planning can exploit predictable price structure but may become stale, while short-horizon planning can react to changing prices, capacity, and progress but lacks lookahead. The scheduler must therefore coordinate decisions across horizons without overcommitting at either level.

\mypar{C3. Scaling joint optimization to provider-scale infrastructure}
Joint optimization over workloads, regions, execution options, and time creates a combinatorial search space that grows with infrastructure scale and workload flexibility. A practical scheduler must solve it efficiently and integrate with existing cluster schedulers.
\section{System Design}
\label{sec:design}
\subsection{Overview}

\begin{figure}[t]
    \centering
    \includegraphics[width=\linewidth]{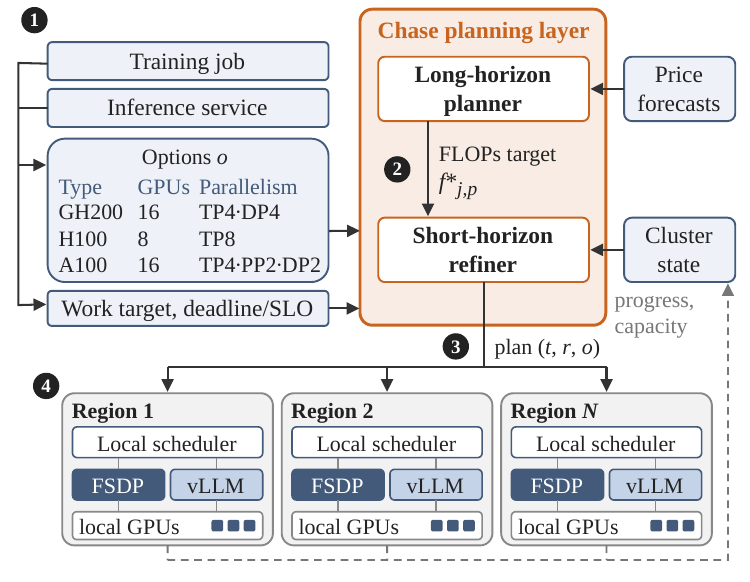}
     \caption{\sys{} as a planning layer. \protect\myc{1} A job enters with its execution options and its scheduling specification. \protect\myc{2} The long-horizon planner sets each deadline-driven job's work per period; the short-horizon refiner binds every job to a step, region, and option. \protect\myc{3} The plan is the contract with the execution layer. \protect\myc{4} Each region's own cluster scheduler and job engines apply the plan to local GPUs and report progress and free capacity back.}
    \label{fig:system-design}
\end{figure}

Three design choices address the challenges of \S\ref{sec:model} (Figure~\ref{fig:system-design}).

\begin{itemize}[itemsep=2pt,topsep=2pt]
\item \textbf{One interface for every workload} (\myc{1}). \sys{} compiles each job into \emph{execution options}, each a resource configuration with its profiled throughput and power draw, and a \emph{scheduling specification} stating what any valid schedule must satisfy. Neither carries an objective, so any two jobs are comparable on energy per unit of work (\S\ref{sec:job-abstraction}).
\item \textbf{A split planning horizon} (\myc{2}). A long-horizon planner commits only how much work each job should complete in each coarse period, and a short-horizon refiner turns that into \emph{spatial-temporal execution plans} naming the option, region, and time step for every job. One work target per job per period is the whole interface between them, so a revision never re-solves the planner (\S\ref{sec:multi-horizon}).
\item \textbf{The plan as the contract with the execution layer} (\myc{3}, \myc{4}). Existing cluster schedulers and job-level engines consume those plans and keep managing local concerns such as queueing and placement. Pre-validated options, option pruning, and the horizon split keep the problem the solver sees small enough to re-solve every slot (\S\ref{sec:system}).
\end{itemize}

\subsection{Job Interface Design}
\label{sec:job-abstraction}

\sys{} compiles every workload into a common scheduling interface, shown in Figure~\ref{fig:system-design}. We use $j$ for a job and $O_j$ for its execution options. An option $o \in O_j$ pairs a resource configuration with profiled behavior: the configuration is a GPU type, a GPU count $W_o$, and a parallelism strategy, and the behavior is a sustained throughput $\Phi_o$ and a power draw $E_o$ (kW).
 
Existing tooling supplies the configuration search and the profiling. Parallelism planners such as Alpa~\cite{zheng2022alpa}, Metis~\cite{um2024metis}, and Sailor~\cite{strati2025sailor} enumerate the GPU counts and parallelism strategies valid for a training job, and serving systems enumerate the configurations that meet an inference service's latency and throughput targets~\cite{agrawal2024vidur}. Each valid configuration is profiled: a short run records its sustained throughput, and an energy profiler reads per-device power through NVML~\cite{nvidia2026nvml}, following the practice of Zeus~\cite{you2023zeus} and MLPerf Power~\cite{tschand2025mlperfpower}. Every option reaching the scheduler is thus feasible by construction, with measured throughput and power.

The scheduling specification states what any valid schedule must satisfy. Two shapes cover every workload we have encountered, and they differ in whether the work requirement is a total or a rate (\S\ref{sec:why-price}). A deadline-driven job states a total FLOPs target $F_j$ and a completion deadline $D_j$. A latency-critical job has no finite $F_j$, because its work arrives continuously; it instead holds a flag to show that it must hold a feasible region at every time step, and its options are profiled at the FLOP/s its offered traffic implies. A service whose demand follows a daily curve enters with options profiled at the rate expected in each period. This is the only point in the design where the scheduler distinguishes job types; \S\ref{sec:practical} adds further constraints without modifying it.

The scheduler's output is a \emph{spatial-temporal execution plan}: a sequence of tuples $(t, r, o)$ giving the time interval, region, and execution option for each job. Because the option already encodes GPU type, count, and parallelism, each tuple tells an existing cluster scheduler~\cite{yoo2003slurm, burns2016borg, narayanan2020gavel} enough to launch the job with no further interaction.

\subsection{A Multi-Horizon Scheduling Mechanism}
\label{sec:multi-horizon}
\begin{figure}[t]
    \centering
    \includegraphics[width=\linewidth]{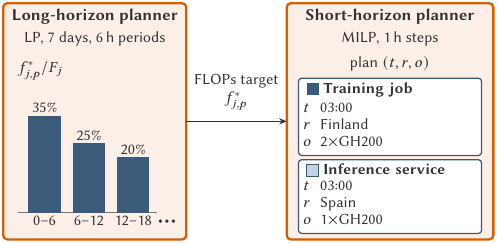}
    \caption{Interface between planners. Bars give the share of one deadline-driven job's total work due in each period, and the arrow carries the current period's target, which is all the refiner requires once the short-term forecast sharpens.}
    \label{fig:horizon-mapping}
\end{figure}

\begin{figure}[t]
    \centering
    \includegraphics[width=\linewidth]{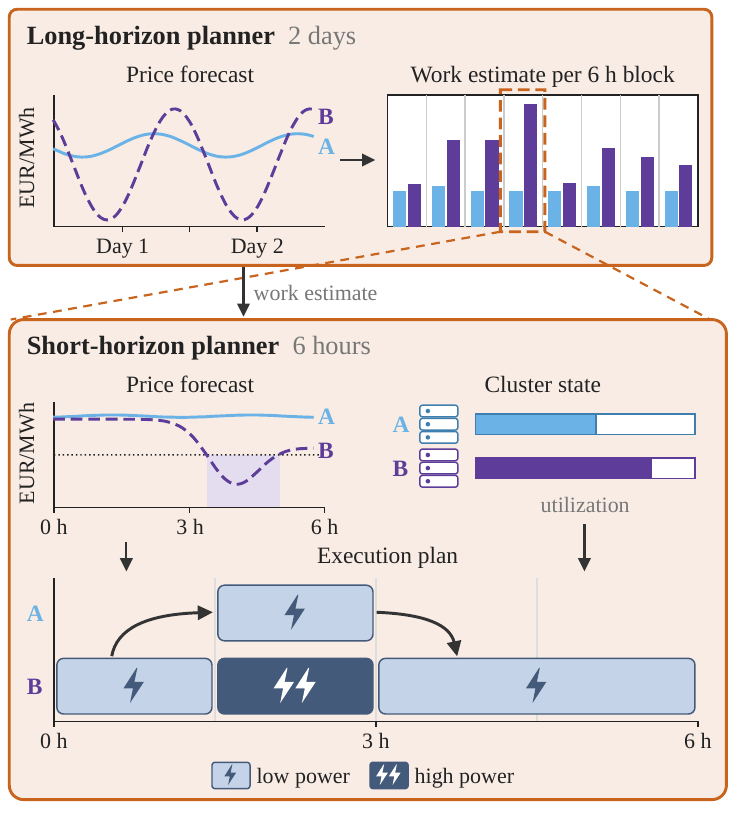}
    \caption{The long-horizon planner reads a multi-day price forecast for every region and estimates how much work each job should complete in each 6~h block. The refiner expands one block, marked by the dashed outline, with a near-term forecast and the current cluster state, resolving it into an execution plan over hour-scale steps.}
    \label{fig:horizon-view}
\end{figure}

The interface makes every workload schedulable under one cross-region scheduler, but leaves open how control is organized across time. \sys{} splits control into a \emph{long-horizon planner} and a \emph{short-horizon refiner} connected through a narrow interface (Figure~\ref{fig:horizon-mapping}). Two properties separate them: the \emph{window} is how far ahead a level looks, and the \emph{granularity} is the length of the time unit it decides over. The planner takes a long window at coarse granularity, reads a price forecast for every region over it, and passes one number per job per period, a work target. The refiner takes a short window at fine granularity, reads that target together with the prices, free capacity, and job progress observed when it runs, and decides which execution option each job runs, in which region, and when. Its output is the plan the regional cluster schedulers receive. Table~\ref{tab:notation} summarizes the notation. Figure \ref{fig:horizon-view} follows one deadline-driven job through both levels.

\subsubsection{Long-Horizon Planning}

The long-horizon planner reasons over a window of several days, divided into coarse periods; we use a 7-day window of 6-hour periods (\S\ref{sec:eval}). The goal is to spread work across regions and periods to capture broad cost structure, not to commit to placements. Because all jobs draw on the same regional capacity, it resolves the contention between them directly, routing power-intensive work to cheap-energy regions.

The planner allocates FLOPs across options, regions, and periods to minimize forecast energy cost. Let $j \in J$ index jobs, $o \in O_j$ options, $r \in R$ regions, and $p \in P$ coarse periods. Jobs are split by the shape of their scheduling specification (\S\ref{sec:job-abstraction}) rather than by what they compute: $J_{\text{dl}}$ carries a total work target and a deadline, $J_{\text{lc}}$ carries a requirement to stay available. The per-FLOP cost $\bar{C}_{r,p} \cdot E_o \Delta t / \Phi_o$ shows why a power-hungry but fast option can be cheap per unit of work.

\begin{align}
\min_{f} \;& \sum_{\substack{j,o,r,p}}
  \bar{C}_{r,p} \cdot \tfrac{E_o \, \Delta t}{\Phi_o} \cdot f_{j,o,r,p} \\
\text{s.t.} \;
& \sum_{\substack{o,r,p}} f_{j,o,r,p} = F_j
  \quad \forall\, j \in J_{\text{dl}}
  \label{eq:flop_total} \\
& \sum_{\substack{o,r}} \tfrac{f_{j,o,r,p}}{\Phi_o} = |T_p|
  \quad \forall\, j \in J_{\text{lc}},\, p
  \label{eq:serve_occupancy} \\
& \sum_{\substack{j,o}} W_o \cdot \tfrac{f_{j,o,r,p}}{\Phi_o} \leq K_r \cdot |T_p|
  \quad \forall\, r,p
  \label{eq:upper_cap} \\
& f_{j,o,r,p} \geq 0
  \quad \forall\, j,o,r,p
\end{align}

\begin{table}[t]
\centering
\caption{Notation used in the multi-horizon formulation.}
\label{tab:notation}
\small
\begin{tabular}{@{}cp{4.8cm}@{}}
\toprule
\multicolumn{2}{c}{\textbf{Sets}} \\
\midrule
$J$ & All jobs; $J_{\text{dl}}$ deadline-driven, $J_{\text{lc}}$ latency-critical \\
$O_j$ & Execution options for job $j$ \\
$R$ & Regions \\
$P$, $T$ & Coarse periods; fine time steps \\
$T_p$ & Fine time steps within period $p$ \\
\midrule
\multicolumn{2}{c}{\textbf{Parameters}} \\
\midrule
$\Phi_o$ & Throughput of option $o$ (FLOP/step) \\
$E_o$ & Power draw of option $o$ (kW) \\
$W_o$ & GPU count of option $o$ \\
$\Delta t$ & Length of one time step (h) \\
$F_j$ / $F_j^{\text{rem}}$ & Total / remaining work of job $j$ (FLOP) \\
$D_j$ & Completion deadline of job $j$ \\
$\bar{C}_{r,p}$ / $C_{r,t}$ & Forecast / actual electricity price \\
$K_r$ & GPU capacity of region $r$ \\
\midrule
\multicolumn{2}{c}{\textbf{Decision variables}} \\
\midrule
$f_{j,o,r,p}$ & Work allocated, FLOP (long-horizon) \\
$x_{j,o,r,t}$ & $\in\{0,1\}$: placement (short-horizon) \\
\midrule
\multicolumn{2}{c}{\textbf{Derived}} \\
\midrule
$f_{j,p}^{*}$ & Work target from long-horizon (FLOP) \\
$\tau_{j,o,r \to r'}$ & Profiled migration stall (h) \\
$y_{j,o,r,r',t}$ & $\in\{0,1\}$: $j$ moves $r \to r'$ at step $t$ \\
$\mathcal{M}(j, x)$ & Migration cost, Eq.~\ref{eq:migration} \\
\bottomrule
\end{tabular}
\end{table}

Eq.~\ref{eq:flop_total} ensures all of a deadline-driven job's work is planned for. Eq.~\ref{eq:serve_occupancy} is the latency-critical counterpart: such a job has no finite work target, so the planner must instead ensure uptime. Because the LP is continuous it may split it across regions within a period; the refiner binds it to one. Eq.~\ref{eq:upper_cap} ensures the implied GPU time does not exceed regional capacity, counting both shapes, so a latency-critical job's occupancy displaces deadline-driven work from a cheap region. The output is a FLOPs target $f_{j,p}^{*}$ per deadline-driven job per period, representing how much work should be completed given the forecast cost landscape. A latency-critical job has no target, since its requirement is the same in every period. That target is the only quantity the planner commits to; the refiner determines the execution option, the region, and the time step under current prices and capacity.

\subsubsection{Short-Horizon Refinement}

The short-horizon refiner turns the coarse targets into concrete assignments over a rolling window of a few hours divided into fine time steps, using the prices, capacity, and job progress it observes; our evaluation uses a 6-hour window of 1-hour steps, re-solved every step (\S\ref{sec:eval}). For deadline-driven jobs it must deliver the FLOPs target $f_{j,p}^{*}$ at minimum energy cost. For latency-critical jobs it enforces placement directly, taking the cheapest feasible region at each step.

The refiner minimizes the sum of compute energy cost and migration cost, where $\mathcal{M}(j,x)$ captures the overhead of switching a job's region or configuration between adjacent time steps. Here $t \in T$ indexes fine time steps within the rolling window, and $p$ is the coarse period it falls in:

\begin{align}
\min_{x} \;& \sum_{\substack{j,o,r,t}}
  C_{r,t} \cdot E_o \, \Delta t \cdot x_{j,o,r,t}
  \;+\; \sum_{j} \mathcal{M}(j, x) \\
\text{s.t.} \;
& \sum_{\substack{o,r,t}} \Phi_o \cdot x_{j,o,r,t}
  \geq f_{j,p}^{*}
  \quad \forall\, j \in J_{\text{dl}}
  \label{eq:flop_lower} \\
& \sum_{\substack{o,r,t}} \Phi_o \cdot x_{j,o,r,t}
  \leq F_j^{\text{rem}}
  \quad \forall\, j \in J_{\text{dl}}
  \label{eq:flop_cap} \\
& \sum_{\substack{j,o}} W_o \cdot x_{j,o,r,t}
  \leq K_{r}
  \quad \forall\, r,t
  \label{eq:capacity} \\
& \sum_{\substack{o,r}} x_{j,o,r,t} \leq 1
  \quad \forall\, j \in J_{\text{dl}},\, t
  \label{eq:one_option} \\
& \sum_{\substack{o,r}} x_{j,o,r,t} = 1
  \quad \forall\, j \in J_{\text{lc}},\, t
  \label{eq:serve_place} \\
& x_{j,o,r,t} \in \{0,1\}
  \quad \forall\, j,o,r,t
\end{align}

The objective balances two costs. Compute cost is the energy price times power draw times step length, paid only while a job runs; substituting carbon intensity for $C_{r,t}$ optimizes emissions instead~\cite{radovanovic2023carbon, hanafy2023carbonscaler, sukprasert2024limitations}. Migration cost penalizes placement changes across time steps, including pausing, resuming, and cross-region moves.

Five constraints govern feasibility. Eq.~\ref{eq:flop_lower} ensures each deadline-driven job meets its FLOPs target from the long-horizon planner. Eq.~\ref{eq:flop_cap} bounds the allocated work by the job's remaining FLOPs, which matters when prices turn negative. Eq.~\ref{eq:capacity} enforces per-region GPU capacity at every time step. Eq.~\ref{eq:one_option} holds a deadline-driven job to at most one option in one region per step. Eq.~\ref{eq:serve_place} is where latency-critical placement is enforced: such a job must hold exactly one option in one region at every step. Its options already satisfy the service targets it was admitted under, so the objective puts it in the cheapest region with sufficient capacity.

\mypar{Migration overhead}
Migration cost is workload-specific and route-specific. Checkpoint-based migration saves model state~\cite{mohan2021checkfreq, wan2025bytecheckpoint}, transfers it across regions~\cite{jain2023skyplane}, and resumes on new hardware. \sys{} treats migration cost as a profiling problem: before submission we measure, for each execution option and each ordered pair of regions, the overhead $\tau_{j,o,r \to r'}$, from the checkpoint in $r$ to the first resumption in $r'$, and update it as network conditions change.

The migration cost must use the same units as the price difference motivating the move. During $\tau_{j,o,r \to r'}$ the job delivers nothing, so $\Phi_o \tau_{j,o,r \to r'}$ FLOPs must be delivered in a later step. At $E_o / \Phi_o$ kWh per FLOP, replacing them costs $E_o \tau_{j,o,r \to r'}$ kWh, charged at the destination region's price:

\begin{equation}
\mathcal{M}(j, x) = \sum_{t} \sum_{o,r,r'}
  C_{r',t} \cdot E_o \cdot \tau_{j,o,r \to r'} \cdot y_{j,o,r,r',t}
\label{eq:migration}
\end{equation}

Here $y_{j,o,r,r',t} = 1$ when job $j$ runs option $o$ and moves from region $r$ at step $t-1$ to $r'$ at step $t$, which the refiner linearizes as $y_{j,o,r,r',t} \geq x_{j,o,r,t-1} + x_{j,o,r',t} - 1$. A pause in place is the same expression with $r = r'$ and the pause-and-resume stall. Written this way a move is taken only when the price gap it opens is worth more than the energy needed to replace the work it costs, which is why \sys{} is less likely to choose migration at low bandwidth (\S\ref{sec:eval:mig}).

\subsection{Tractable Scheduling at Scale}
\label{sec:system}

The formulation above abstracts away the cost of solving it; two design choices bound that cost. The unified abstraction (\S\ref{sec:job-abstraction}) admits a job as a finite set of pre-validated options, so $|O_j|$ is fixed before the solver runs rather than searched over. The horizon split bounds $|T|$: the planner stays a lightweight LP over coarse blocks, and the refiner solves its MILP over a bounded rolling window rather than the full horizon.

\mypar{Option pruning}
\sys{} ranks each job's options by energy efficiency, $\Phi_o / E_o$ in FLOP per kWh, and passes only the top $k$ to the refiner. Delivering one FLOP through option $o$ at price $C_{r,t}$ costs $C_{r,t} \cdot E_o / \Phi_o$, so the ordering is the same in every region and every step, and dropping a less efficient option cannot change the optimum while no capacity constraint binds. Efficiency is unrelated to size or absolute speed, so a job's most efficient option may ask for 16 GPUs where 8 are free, or run too slowly for a deadline a less efficient option would meet. Keeping the top $k$ leaves the refiner a choice of GPU count and keeps it solvable within seconds at scale (\S\ref{sec:eval}).

\subsection{Additional Practical Considerations}
\label{sec:practical}

\mypar{Execution requirements}
Carrying out a plan requires pause, checkpoint, migrate, and resume from job engine. PyTorch FSDP~\cite{zhao2023fsdp}, HuggingFace Transformers~\cite{wolf2020transformers}, and vLLM~\cite{kwon2023vllm} already provide them, so \sys{} delegates execution (\S\ref{sec:impl}). 

\mypar{Workload-specific constraints}
Beyond Eq.~\ref{eq:flop_lower}--\ref{eq:serve_place}, the refiner carries a deadline constraint: a deadline-driven job must complete its total FLOPs target by $D_j$, not only the per-window quota, which binds when a job has fallen behind. When capacity or a progress deviation leaves no feasible assignment, the refiner evicts the least critical jobs and re-solves, and the planner absorbs the deviation at its next cycle.

\mypar{Changing job behavior}
Execution options may become stale as workloads evolve. The profiling layer then updates the options and their throughput and power profiles, and \sys{} uses them from the next scheduling window.
For post-training jobs of uncertain duration, such as reinforcement learning ~\cite{shao2024deepseekmath, sheng2025hybridflow}, \sys{} uses the maximum estimated FLOPs and allows early exit.

\mypar{Plan stability}
\sys{} avoids unnecessary migration by accepting a new plan only if energy cost is lowered by more than a threshold.

\section{Implementation}
\label{sec:impl}

We implement \sys{} prototype on Slurm with one cluster per region~\cite{yoo2003slurm}, and a central controller issues Slurm commands to each cluster. Regions and their GPU capacity are declared in a JSON configuration. Each job is submitted with a JSON specification of its workload, data, and execution options. Inference specifications also list the regions that meet the service's latency target. In our prototype, training data is replicated in every region. Where full replication is impractical, data can be moved with existing inter-region bulk transfer systems~\cite{jain2023skyplane} or streamed from remote object storage while training proceeds~\cite{mosaicml2026streaming}.

\mypar{Plan execution}
The launcher reads the GPU count and parallelism layout of its assigned option. To preempt a training job, \sys{} issues \texttt{scancel -{}-signal=USR1}, the launcher writes a PyTorch DCP checkpoint~\cite{pytorch2026dcp}, and \texttt{scontrol requeuehold} releases its GPUs. \texttt{scontrol release} resumes the job at its planned start. Resizing updates the held job via \texttt{scontrol update NumNodes= TresPerNode=gres/gpu:}, and DCP reshards the checkpoint on load. Migration transfers the checkpoint and job state over the network to the destination cluster and resubmits the job with \texttt{sbatch -M <cluster>}. Inference is scheduled per serving instance, and each instance runs as a Slurm job. To relocate an instance, \sys{} submits a replacement job to the destination cluster with \texttt{sbatch -M <cluster>}. The serving system's router begins routing requests to the new instance once it passes its readiness check and stops routing to the source, after which \sys{} cancels the source job with \texttt{scancel}.

\mypar{Monitoring and correction}
Every 30\,s, each job reports a heartbeat containing its progress, GPU power, and, for inference, request rate. Training progress is the step count the launcher records in the job's \texttt{Comment} field, and the reported power is measured by Slurm's NVML-based \texttt{acct\_gather\_energy/gpu} plugin~\cite{nvidia2026nvml}. Free capacity is obtained from \texttt{sinfo}~\cite{schedmd2026sinfo} and prices from the ENTSO-E day-ahead API~\cite{entsoe2026transparency}. \sys{} corrects two sources of deviation from the plan. Work shortfalls caused by delayed or failed actions or lost capacity are added to the job's work requirement for the next period. Price forecasts are adjusted by an exponential moving average of past forecast error that weights underestimated periods more heavily, since these directly reduce savings.

\mypar{Solver}
Both programs are solved with CPLEX~\cite{ibm2026cplex} over each job's top $k$ options by $\Phi_o / E_o$. A new plan is adopted only if it clears the stability threshold of \S\ref{sec:practical}.

\mypar{Portability}
The system is extendable to other control planes. On Kubernetes, device plugins and Kueue~\cite{kubernetes2026deviceplugins,kubernetes2026kueue} expose GPU inventory and placement constraints, Kubeflow Trainer with NFS-backed persistent volumes~\cite{kubeflow2026trainer,kubernetes2026pv} supports checkpoint-based preemption and restart, and the Kubernetes API~\cite{kubernetes2026api} and DCGM~\cite{nvidia2026dcgm} provide workload state and power telemetry. The planner requires no modification.
\section{Evaluation}
\label{sec:eval}

We evaluate \sys{} on real hardware and in an emulated multi-region deployment. The evaluation addresses the challenges introduced in \S\ref{sec:model}:

\begin{enumerate}[leftmargin=*,itemsep=0pt,topsep=1pt]
\item \sys outperforms all baselines on three price traces, saving 27--35\% on volatile markets, 12\% on stable markets, and up to 249\% under negative prices, where compute generates revenue (\S\ref{sec:eval:e1});
\item joint scheduling through the job abstraction eliminates electricity cost on the training-heavy mix and yields net revenue on the other mixes (\S\ref{sec:eval:e2});
\item multi-horizon planning outperforms single-horizon alternatives, and self-correction limits the impact of forecast error and execution variability (\S\ref{sec:eval:horizon}, \S\ref{sec:eval:e3});
\item the planning layer scales to 200 jobs across 39 regions and over 40{,}000 GPUs, solving fine-grained placement in under 25 seconds (\S\ref{sec:eval:e4}).
\end{enumerate}

\subsection{Setup}

We evaluate at two scales: a \emph{physical deployment} on real GPUs to validate end-to-end performance and correctness, and an \emph{emulated deployment} for policy quality, ablations, and scalability beyond what the testbed permits.

\mypar{Physical testbed}
We deploy \sys{} on four 4-GPU NVIDIA GH200~\cite{nvidia2026gh200} nodes, split into two regions replaying the Germany and Finland traces. (\S\ref{sec:eval:detail}).

\mypar{Profiling}
We profile each (model, GPU count, parallelism strategy) tuple for steady-state throughput (FLOP/s) and power draw (W), sampling device power through NVML~\cite{nvidia2026nvml}.
We also profile migration cost for all models assuming cross-region checkpoint transfer bandwidth of ${\sim}$10\,Gbps, measuring the end-to-end overhead.
Inference options are profiled at saturation, the request rate at which the configuration just meets its latency target, so every inference job is charged the highest power it can draw. This is the worst case for us: a server carrying less traffic draws less and frees headroom, which \sys{} would see through the same two profiled numbers (\S\ref{sec:job-abstraction}).
These profiles define the execution options available to \sys, and all emulated experiments use them, so the emulation reflects real hardware.

\mypar{Emulated deployment}
Scaling to multiple regions is prohibitively expensive on physical hardware, so for all breakdown analyses we run the full system with GPU execution replaced by CPU processes that sleep for each configuration's profiled duration, as LLM inference simulators do~\cite{agrawal2024vidur}.
The default emulated cluster spans 4 European regions (Finland, France, Germany, Spain) with 32 GH200 GPUs each.

\mypar{Electricity prices}
We replay three 7-day traces of day-ahead prices from the ENTSO-E Transparency Platform~\cite{entsoe2026transparency}. \emph{Stable} covers a period with no sharp price moves and no negative prices. \emph{Rotation} covers a period in which the cheapest region rotates. \emph{Negative} covers a period in which 20\% of hours have negative prices from renewable oversupply~\cite{fanone2013negative, seel2021negative}.

\mypar{Workloads}
We create job mixes from three models: GPT-OSS-120B~\cite{openai2025gptoss}, Nemotron-3-Nano-30B-A3B~\cite{nvidia2025nemotron}, and Qwen3.5-35B-A3B~\cite{qwen2026qwen35}.
Training jobs are a mix of small (${\sim}$50\,GPU-hours) and large (${\sim}$500\,GPU-hours), and expose multiple pre-validated execution options over different GPU configurations and parallelism strategies.
Varying the job count sets saturation, the share of cluster capacity the offered load occupies: the default mix loads 40\% (35 jobs) and a heavier mix 70\% (65 jobs).
The two levels bracket the regimes that decide how much arbitrage is available. At 40\% the cluster has headroom the scheduler can rearrange work into; at 70\% most cheap slots are already contended and every scheduler is forced toward the same placements (\S\ref{sec:eval:e1}). Reporting both separates savings that come from the policy from savings that come from an empty cluster.

\mypar{Scheduler configuration}
Unless stated otherwise, \sys{} runs a 7-day long-horizon window of 6-hour coarse periods and a 6-hour short-horizon rolling window of 1-hour time steps, with option pruning at $k = 3$.
The refiner re-solves every 1-hour step and the planner once per 6-hour period, so a slot in the rest of this section is one hour, and the baselines decide on the same boundaries.

\mypar{Baselines}
We compare against three baselines that expose different forms of scheduling flexibility.
\begin{itemize}[itemsep=0pt,topsep=1pt]
\item \textbf{Static} sends each job to the cheapest region with free capacity at submission and never revisits it, so it adapts neither in time nor in space. The figures label it MAST~\cite{choudhury2024mast}.
\item \textbf{GREEN}~\cite{xu2025green} shifts jobs in time within one region. We extend it to support both training and inference.
\item \textbf{Caribou}~\cite{gsteiger2024caribou} is a carbon-aware planner for serverless workflows, which it places stage by stage, offloading the stages that are not on the critical path. We adapt its solver from spatial assignment over DAG stages to spatial-temporal placement over time slices for all jobs, matching the decision space of \sys{}. The adaptation keeps its per-job scope: it plans each job against the price signal alone.
\end{itemize}
The end-to-end comparison and the workload-mix study report all three. The hardware and migration sweeps (\S\ref{sec:eval:hw}, \S\ref{sec:eval:mig}) vary a parameter of \sys{}'s own mechanism rather than of the workload, so they report savings against Static alone, which keeps the reference fixed.

\mypar{Metrics}
Our primary metric is \emph{cost per completed job} across all jobs, normalized to the most expensive scheduler on each trace.
Deadlines are a feasibility constraint rather than an objective, so we report deadline misses only where a variant fails to meet them (\S\ref{sec:eval:horizon}, \S\ref{sec:eval:e3}).

\subsection{End-to-End Results on Physical Hardware}
\label{sec:eval:detail}

\begin{figure}[t]
  \centering
  \includegraphics[width=\columnwidth]{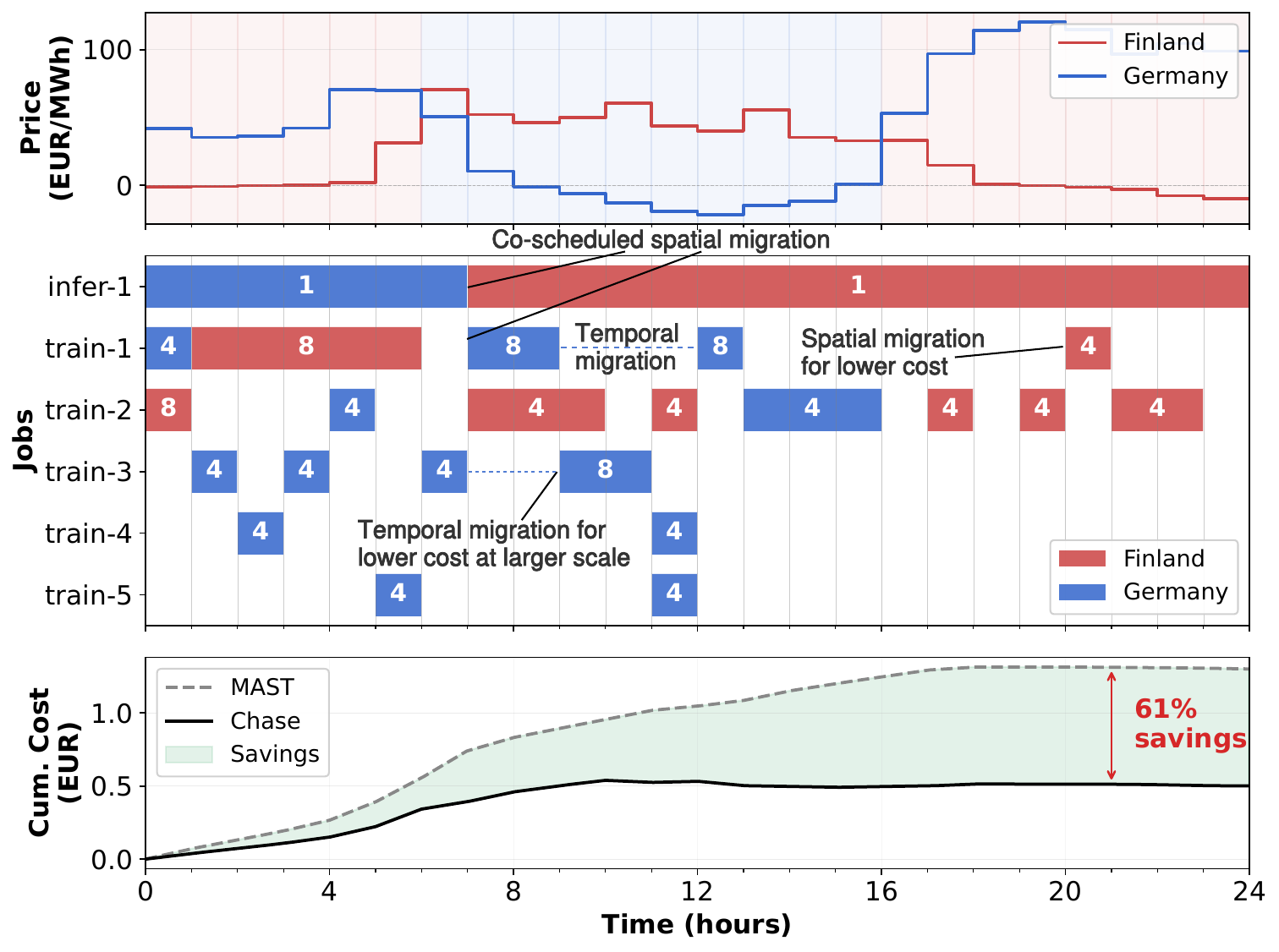}
  \caption{Execution on the physical testbed with two emulated regions. \textbf{Top:} prices for Finland and Germany. \textbf{Middle:} \sys's per-job placement decisions (region and GPU count per slot). \textbf{Bottom:} cumulative cost, \sys{} vs.\ Static.}
  \label{fig:execution_detail}
\end{figure}

\begin{figure*}[t]
  \centering
  \includegraphics[width=.95\textwidth]{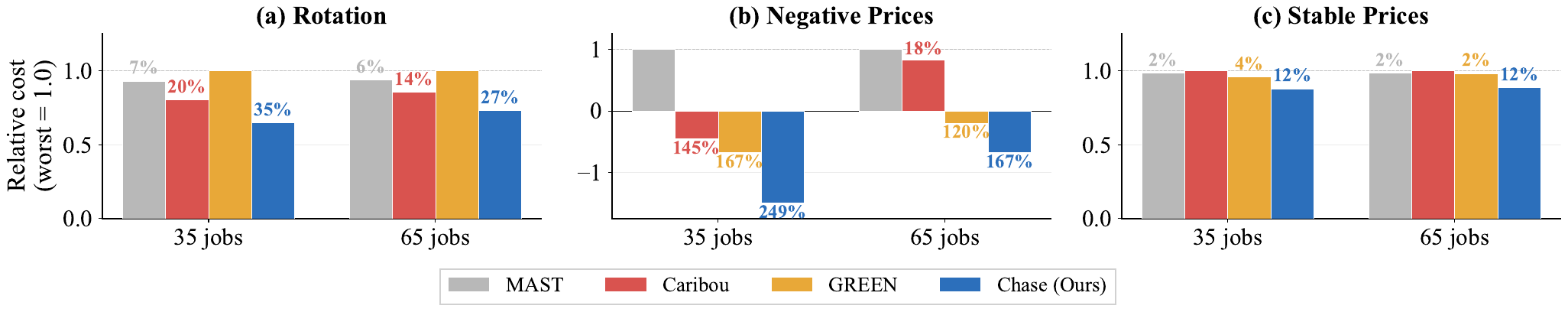}
  \caption{Normalized cost per completed job across three price traces at two saturation levels, 40\% and 70\% (labeled by job count), normalized to the worst scheduler per trace. Bars below zero indicate net profit.}
  \label{fig:comparison}
\end{figure*}

Figure~\ref{fig:execution_detail} shows a 24-hour execution of \sys{} on the physical testbed, with six jobs across two regions whose prices diverge by over 100\,EUR/MWh at their widest.

\mypar{Spatial shifting}
The inference service (infer-1) starts in Germany and moves to Finland at hour 7.
It does not follow the cheapest price: Germany turns cheap at that hour, but \sys gives that capacity to the 8-GPU train-1 job and pushes inference to Finland, an exchange the middle panel marks.
Training jobs migrate when the two prices cross, as train-2 does repeatedly and train-1 at hour 20.

\mypar{Temporal shifting}
Rather than spreading training evenly, the solver concentrates execution in cheap windows and idles jobs in expensive ones.
train-1 pauses after hour 9 and resumes at hour 12 inside the German negative-price window, and train-3 pauses and resumes at a larger GPU count in the same window, both marked in the figure.

\mypar{Selecting execution options}
When prices are low, the solver selects pre-validated options with larger GPU allocations (4 to 8 GPUs), accelerating progress while electricity is cheap, and smaller ones when prices rise.

The bottom panel shows the cumulative effect: \sys accumulates cost slowly during cheap windows, while Static pays the prevailing rate throughout.
The result is a 61\% reduction in total electricity cost over 24 hours.

\subsection{End-to-End Cost Comparison}
\label{sec:eval:e1}

We compare \sys against the baselines on three real price traces, each stressing a different aspect of the problem.

\mypar{Rotation}
When the cheapest region rotates among four countries (Figure~\ref{fig:comparison}(a)), GREEN is the most expensive scheduler: it pins each job to one region at submission and shifts only in time, missing the spatial opportunities rotation creates.
At 40\% saturation, \sys reduces cost by 35\% relative to GREEN, compared to 20\% for Caribou and 7\% for Static.
The gap narrows at 70\% as the heavier load leaves less capacity to rearrange, but \sys remains the cheapest scheduler at both saturation levels. Caribou captures some spatial opportunity, but less than joint optimization across all jobs.

\mypar{Negative prices}
In Figure~\ref{fig:comparison}(b), every scheduler except Static reaches negative total cost at 40\% saturation, meaning it \emph{profits} from negative price windows.
\sys{} exploits them most aggressively: it reduces cost by 249\% relative to Static by placing the most power-consuming jobs into those periods.
Caribou and GREEN also profit (145\% and 167\% respectively) but capture less, because they do not coordinate across jobs.
At 70\% the advantage persists (167\% for \sys vs.\ 120\% for GREEN and 18\% for Caribou), though the higher load reduces the scheduling headroom.

\mypar{Stable prices}
This trace is the limiting case for \sys{}: prices are stable across regions and time with no negative hours, leaving little room for spatial or temporal arbitrage (Figure~\ref{fig:comparison}(c)).
\sys still achieves the lowest cost, saving 12\% over the most expensive baseline (Caribou) in both cases, while Static and GREEN save 2--4\%.
Caribou falls behind even Static because it plans one job at a time: with so narrow a spread, independent choices converge on the same few regions, and jobs planned later fall back to worse placements. Static at least spreads load, since it only takes regions with free capacity, which on this trace costs less than crowding.

\begin{figure*}[t]
  \centering
  \begin{subfigure}[b]{0.31\textwidth}
    \centering
    \includegraphics[width=\textwidth]{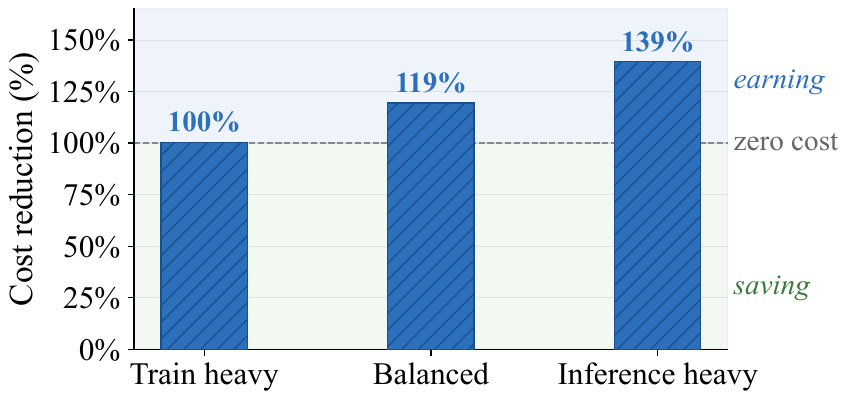}
    \caption{Common job abstraction.}
    \label{fig:joint-scheduling}
  \end{subfigure}
  \hfill
  \begin{subfigure}[b]{0.31\textwidth}
    \centering
    \includegraphics[width=\textwidth]{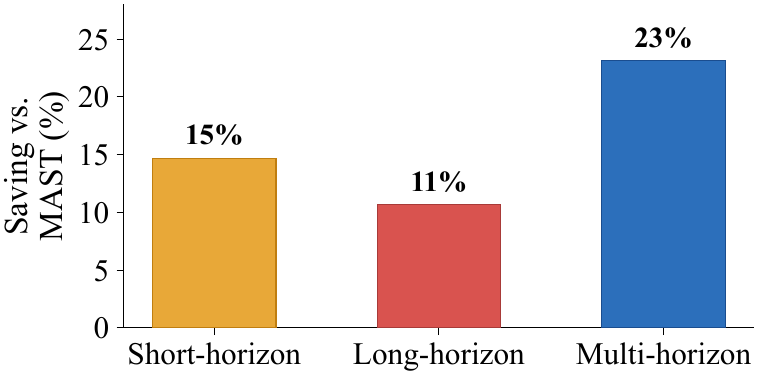}
    \caption{Multi-horizon ablation.}
    \label{fig:horizon}
  \end{subfigure}
  \hfill
  \begin{subfigure}[b]{0.31\textwidth}
    \centering
    \includegraphics[width=\textwidth]{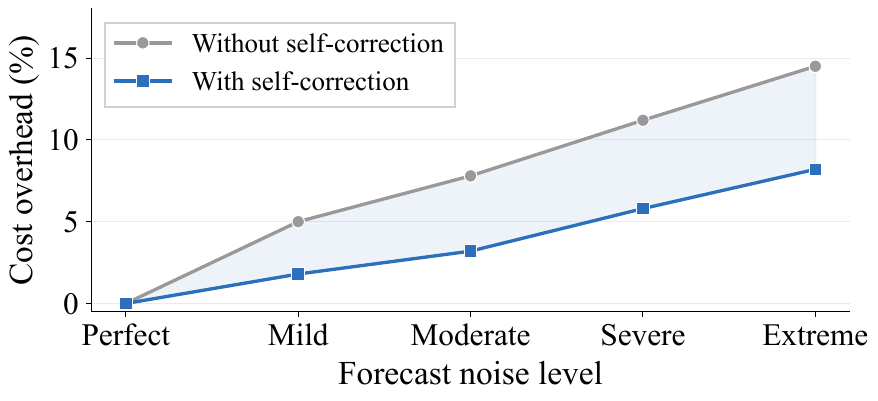}
    \caption{Forecast noise robustness.}
    \label{fig:self-correction}
  \end{subfigure}
  \caption{\textbf{(a)}~Cost reduction of joint over per-class scheduling; 100\% denotes zero cost and values above 100\% denote net revenue. \textbf{(b)}~Savings over Static for single-horizon ablations on the 33 common completions. \textbf{(c)}~Cost overhead relative to the perfect-forecast oracle as noise increases.}
\end{figure*}

\subsection{Breakdown Analysis}

\subsubsection{Value of the Common Job Abstraction}
\label{sec:eval:e2}

We compare \sys's single formulation against a \emph{per-class} baseline that gives each class its own solve.
It runs \sys's planner and refiner twice over the same cluster: once for the inference jobs on the full capacity, then once for the training and post-training jobs on whatever the first solve left, without revising the first.
The trace, options, capacity, and solver are identical; the only difference is whether the two classes are ranked together.
Figure~\ref{fig:joint-scheduling} reports results for three mixes: \emph{training heavy}, \emph{balanced}, and \emph{inference heavy}.

One solver for both types eliminates electricity cost entirely on the training-heavy mix (100\% reduction), and turns it into net revenue on the balanced (119\%) and inference-heavy (139\%) mixes.
This is the value of the common job abstraction: every workload exposes what global placement needs through the same interface, so \sys ranks heterogeneous jobs for the same capacity and moves lower-power jobs to moderately priced regions in the hours when higher-power jobs gain most. A scheduler that reasons about one class at a time cannot make this allocation, because whichever class it plans first claims the cheap windows.

\subsubsection{Value of Multi-Horizon Planning}
\label{sec:eval:horizon}

\begin{table}[t]
\centering
\caption{Cost reduction over Static as \sys features are enabled. The last
three rows are alternatives to each other, each built on top of joint
scheduling.}
\label{tab:design-insight-comparison}
\footnotesize
\begin{tabular}{lr}
\toprule
\textbf{Policy} & \textbf{Cost reduction (\%)} \\
\midrule
Static                        & ---  \\
+ Migration                   &  27  \\
+ Joint scheduling            &  54  \\
\quad with short horizon only  &  59  \\
\quad with long horizon only   &  71  \\
\quad with both horizons       &  79  \\
\bottomrule
\end{tabular}
\end{table}

We isolate multi-horizon planning with two ablations: \emph{short horizon only} runs the refiner over successive 6-hour windows without coarse guidance, and \emph{long horizon only} runs the planner over the full horizon but fills the fine-grained time slots within each period greedily.
The long-horizon-only variant assigns more work to some coarse periods than the greedy fine-grained scheduler can realize, so 2 of the 35 jobs do not finish by their deadline. We therefore compare cost on the 33 jobs that every variant completes, since a variant that drops work would otherwise look cheaper for the wrong reason.

In Figure~\ref{fig:horizon}, the full system saves 23\% over Static, compared with 15\% and 11\% for the short- and long-horizon-only ablations, so removing either component costs 8--12 percentage points.

We then enable features incrementally from Static on the rotation trace (Table~\ref{tab:design-insight-comparison}), in a 24-hour scenario with a fine-tuning job and an inference service across three regions.
Migration alone saves 27\%, but letting one solver rank both jobs doubles that, because it can give the scarce cheap capacity to the higher-power job rather than to whichever was planned first, and the full combination reaches 79\%.

\subsubsection{Self-Correction}
\label{sec:eval:e3}

\sys{} works from day-ahead forecasts that carry error, and job runtimes deviate from profiles because of stragglers. We evaluate self-correction under both.

\mypar{Price forecast noise}
We add Gaussian noise to the price data and compare five noise levels from \emph{perfect} (0\% MAPE) to \emph{extreme} (${\sim}$60\% MAPE).
Typical day-ahead forecasts fall between our mild and moderate settings (5--15\% MAPE), consistent with state-of-the-art forecasting accuracy~\cite{lago2021forecasting}.

Figure~\ref{fig:self-correction} reports cost relative to a perfect-information oracle.
Without self-correction, cost rises to +11.2\% at severe (46\% MAPE) and +14.5\% at extreme; self-correction roughly halves the penalty at every level, to 5.8\% and 8.2\%.

\mypar{Runtime stragglers}
Throughput deviations are common at cluster scale~\cite{kokolis2025reliability, lin2025stragglers}, so job progress deviates from profiles.
We perturb job throughput independently of price: each slot, every active job has a 10\% probability of running at 0.3--0.8$\times$ its profiled speed, which raises cost by 4.1\% and makes 2 additional jobs miss their deadline.
The scheduler treats stragglers as behind-schedule jobs and compensates in later slots, requiring no mechanism beyond reoptimization.

\subsubsection{Workload Mix Sensitivity}

\begin{table}[t]
\centering
\caption{Cost reduction (\%) of \sys{} vs.\ baselines on the rotation trace. \textbf{Top:} training job size mix at a fixed 20 training + 10 inference jobs; L, M, S are 16-, 8-, and 4-GPU jobs. \textbf{Bottom:} train:infer composition at a fixed job count.}
\label{tab:mix_sensitivity}
\footnotesize
\begin{tabular}{@{}lccc@{}}
\toprule
Mix & vs.\ Static & vs.\ GREEN & vs.\ Caribou \\
\midrule
\multicolumn{4}{@{}l}{\emph{Training job size}} \\
20M                & 11.2 & 25.2 &  9.4 \\
10M + 10S          & 35.5 & 28.1 & 13.0 \\
10L + 10S          & 15.2 & 24.9 &  4.2 \\
3L + 7M + 10S      & 16.3 & 29.4 & 12.3 \\
\midrule
\multicolumn{4}{@{}l}{\emph{Composition (train:infer)}} \\
Training-heavy \hfill 80:20  & 30.0 & 27.3 & 19.0 \\
Balanced \hfill 50:50        & 26.5 & 30.0 & 17.3 \\
Inference-heavy \hfill 20:80 & 16.9 & 31.1 & 12.1 \\
\bottomrule
\end{tabular}
\end{table}

\mypar{Training job size distribution}
We fix the job count at 20 training + 10 inference but vary the ratio of training jobs across three GPU counts: L (16~GPU), M (8~GPU), and S (4~GPU).
The top half of Table~\ref{tab:mix_sensitivity} shows that across all mixes, \sys consistently achieves 4--35\% cost reduction over the baselines.

\mypar{Inference vs.\ training ratio}
We vary the fraction of online inference jobs while keeping the total job count the same.
The bottom half of Table~\ref{tab:mix_sensitivity} shows that the reduction against Static and Caribou grows as the training fraction grows, because training is what gives \sys temporal flexibility to exploit.
Compared to GREEN, the temporal-only baseline, \sys consistently achieves around 30\% reduction by arbitraging across the cheapest regions.

\begin{figure}[t]
  \centering
  \begin{subfigure}[b]{0.49\linewidth}
    \centering
    \includegraphics[width=\linewidth]{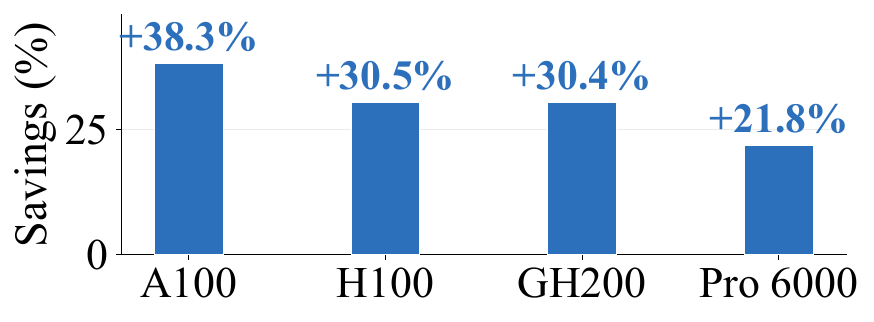}
    \caption{Hardware sensitivity}
    \label{fig:hardware-sensitivity}
  \end{subfigure}
  \hfill
  \begin{subfigure}[b]{0.49\linewidth}
    \centering
    \includegraphics[width=\linewidth]{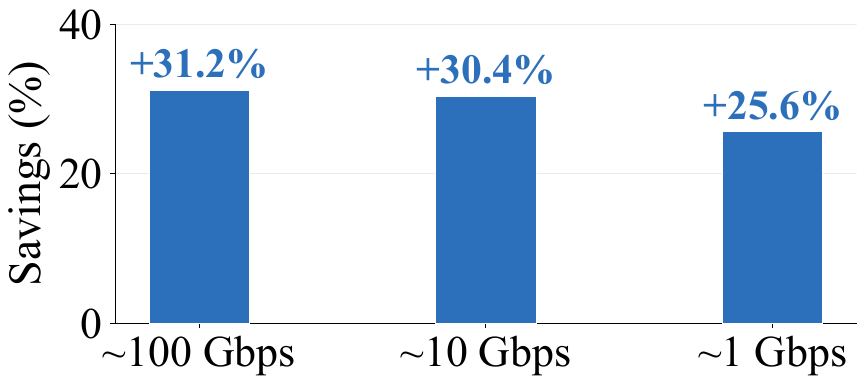}
    \caption{Migration sensitivity}
    \label{fig:migration-sensitivity}
  \end{subfigure}
  \caption{\textbf{(a)}~Savings compared to Static on the rotation trace across four GPU types.
  \textbf{(b)}~Effect of interconnect bandwidth on \sys's savings over Static.
  }
\end{figure}

\subsubsection{Hardware Sensitivity}
\label{sec:eval:hw}

Our testbed holds only GH200 nodes, so we vary GPU type in emulation to see how compute capability affects \sys{}'s advantage: the workload, job count, per-option GPU count, and parallelism strategy remain as in the default mix; only per-GPU compute and power change.
As \autoref{fig:hardware-sensitivity} shows, \sys outperforms Static on the rotation trace across all four GPU types: the A100~\cite{nvidia2026a100}, H100~\cite{nvidia2026h100}, GH200~\cite{nvidia2026gh200}, and the workstation-class RTX PRO 6000~\cite{nvidia2026rtxpro6000}.
Savings range from 21.8\% on the RTX PRO 6000 to 38.3\% on the A100.
The execution options carry hardware-specific performance and power, so the formulation absorbs new GPU types without change.
Savings vary with the work a GPU delivers per unit of energy (FLOPs/kWh).
The A100 delivers the least, so the same job draws the most energy there and puts the most cost in reach of price variation.
At the other end, the lower throughput of the RTX PRO 6000 keeps a job resident for more hours, leaving fewer hours to skip.

\subsubsection{Migration Overhead Sensitivity}
\label{sec:eval:mig}

The main overhead of global migration is WAN-limited transfer latency. \autoref{fig:migration-sensitivity} shows that \sys{} saves over 25\% against Static at every tested bandwidth.
At ${\sim}$100\,Gbps migration is cheap and the solver migrates 15 times over the horizon, saving 31.2\% through combined spatial and temporal shifting.
At ${\sim}$10\,Gbps it migrates more conservatively and still saves 30.4\%, and at ${\sim}$1\,Gbps it finds migration too expensive and forgoes it, still saving 25.6\% through temporal shifting alone.

\subsubsection{Formulation Scalability}
\label{sec:eval:e4}

\begin{table}[t]
  \centering
  \footnotesize
  \caption{Fine MILP solve time across cluster configurations; it scales with jobs\,$\times$\,regions, not with GPUs per region.}
  \label{tab:scalability}
  \setlength{\tabcolsep}{4pt}
  \begin{tabular*}{\linewidth}{@{\extracolsep{\fill}}lrrrr@{}}
    \toprule
    Scenario & Jobs & Regions & Total GPUs & Solve time (s) \\
    \midrule
    AWS, small         &  50 & 39 & 10,296 &  3.8 \\
    AWS, medium        & 100 & 39 & 10,296 & 11.6 \\
    AWS, large         & 200 & 39 & 41,184 & 24.7 \\
    \midrule
    Azure, small       &  50 & 70 & 33,600 & 16.1 \\
    Azure, medium      & 100 & 70 & 33,600 & 59.1 \\
    \bottomrule
  \end{tabular*}
\end{table}

We test whether the formulation remains tractable at provider scale. The coarse LP's solve time is negligible. The fine MILP has a binary variable per (job, option, region, slot) tuple, so solve time follows the product of the job and region counts. We stress-test it on configurations modeled after the published region counts of major cloud providers~\cite{aws2026regions, azure2026regions}, varying jobs from 50 to 200 and regions from 39 to 70.

Table~\ref{tab:scalability} yields two observations.
First, the global formulation scales with the number of scheduling decisions rather than the number of physical GPUs: raising the region count 1.8$\times$ raises solve time 4--5$\times$, while the large AWS configuration has 4$\times$ the GPUs of the medium one but only 2$\times$ its solve time, matching its doubled job count.
Second, every configuration stays well within the one-hour re-solve budget, so the planning layer is not a bottleneck at provider scale.

\section{Related Work}
\label{sec:related-work}

\mypar{GPU cluster scheduling}
Single-cluster GPU schedulers optimize utilization, fairness, or completion
time~\cite{xiao2018gandiva, narayanan2020gavel, qiao2021pollux}, with
Zeus~\cite{you2023zeus} also targeting energy efficiency.
Multi-cluster systems such as MAST~\cite{choudhury2024mast} and
Singularity~\cite{shukla2022singularity} span data centers
for capacity and reliability, but do not jointly optimize electricity price
across regions and time.

\mypar{Geo-distributed training and serving}
Sailor~\cite{strati2025sailor} and decentralized training~\cite{yuan2022decentralized}
exploit geographically distributed resources, while preemptible-capacity
systems~\cite{thorpe2023bamboo, duan2024parcae} handle interruption and recovery.
SkyPilot~\cite{yang2023skypilot}, Skyplane~\cite{jain2023skyplane},
SkyServe~\cite{mao2025skyserve}, and SkyWalker~\cite{xia2025skywalker}
exploit geographic placement or routing for individual workloads. \sys{} coordinates multiple workloads competing for the same regional capacity, exploiting spatial and temporal flexibility together.

\mypar{Energy- and carbon-aware scheduling}
Energy-aware scheduling has long exploited temporal and geographic variation
in electricity supply~\cite{qureshi2009cutting}. Carbon-aware systems shift
work in time~\cite{wiesner2021letswait, radovanovic2023carbon}, across
locations~\cite{zheng2020curtailment}, or across resource configurations~\cite{hanafy2023carbonscaler}
; other work plans energy infrastructure and
capacity~\cite{souza2023ecovisor, acun2023carbon, sun2024skybox}.
Caribou~\cite{gsteiger2024caribou} places serverless workflow stages across
regions, while GREEN~\cite{xu2025green} and GreenFlow~\cite{gu2025greenflow}
shift or resize deep-learning jobs within a cluster.
SustainDC~\cite{naug2024sustaindc} models a data center's compute, cooling, and grid coupling as a control environment, and DCcluster-Opt~\cite{guillenperez2025dcclusteropt} extends that to geo-distributed clusters driven by real price, carbon, and weather traces. Both make it possible to test a policy like ours beyond the scale of any single testbed. \sys{} differs in jointly minimizing electricity cost across heterogeneous workloads, regions, and time while accounting for shared capacity and workload constraints.

\mypar{Energy efficiency of AI systems}
Another line of work reduces the energy required by individual workloads through measurement~\cite{strubell2019energy, patterson2021carbon, tschand2025mlperfpower}, training optimizations~\cite{chung2024perseus}, power management~\cite{qiu2024muserve, patel2024power, stojkovic2025tapas}, and more efficient training and serving engines~\cite{kwon2023vllm, zheng2024sglang, zhong2024distserve, sun2024llumnix, zheng2022alpa, zhao2023fsdp, jiang2024megascale, um2024metis}, while checkpointing systems~\cite{mohan2021checkfreq, wan2025bytecheckpoint} reduce migration cost. These techniques improve the execution options that \sys{} schedules.

\section{Discussion and Limitations}
\label{sec:discussion}

\mypar{Effect on electricity prices} We treat electricity prices as external and the load shift small relative to regional demand. That may not hold as AI infrastructure grows toward gigawatt scale, where concentrated load shifting could raise prices in low-cost periods and narrow the differences \sys exploits, and future schedulers may need to account for their own effect on electricity markets~\cite{li2024transients}.

\mypar{Data residency and regulatory constraints}
Spatial shifting assumes that workloads can move freely between regions, but regulations such as the GDPR~\cite{eu2016gdpr} restrict where certain jobs may execute. \sys{} lets operators specify a per-job set of eligible regions, which the solver respects. Fewer eligible regions mean less spatial arbitrage, and a job pinned to one region leaves \sys{} only temporal shifting, so our savings are an upper bound under strict placement constraints.

\mypar{Trading performance for energy}
\sys{} treats deadlines and service targets as hard constraints, so all reported savings preserve performance (\S\ref{sec:why-price}). Allowing bounded performance degradation would provide additional scheduling flexibility, e.g., by permitting limited deadline extensions or off-peak latency relaxation. This requires only modifying the deadline constraint in \S\ref{sec:practical}; we leave this tradeoff to future work.

\section{Conclusion}

\sys is a global planner that schedules workloads across regions and time to minimize energy cost. \sys{} handles heterogeneous workloads through a common abstraction, splits temporal reasoning into a long-horizon planner that captures structural price trends and a short-horizon refiner that produces executable assignments under current conditions, and integrates with existing schedulers. We believe this scheduling model is a practical foundation for energy- and carbon-aware AI infrastructure.

\clearpage
\bibliographystyle{ACM-Reference-Format}
\bibliography{references}

\end{document}